# A comprehensive catalogue of OB cluster candidates in M31 and their association with giant molecular clouds

Yuan Liang,[1] Guang-Wei Li [2]★, Chao-Wei Tsai [1,2,3]★ and Jingwen Wu [1,2]★

[1]*School of Astronomy and Space Science, University of Chinese Academy of Sciences, Beijing 100049, P. R. China*
[2]*National Astronomical Observatories, Chinese Academy of Sciences, Datun Road A20, Beijing, P. R. China*
[3]*Institute for Frontiers in Astronomy and Astrophysics, Beijing Normal University, Beijing 102206, P. R. China*

## ABSTRACT

We present a new catalogue of 578 OB cluster (OBC) candidates in the Andromeda galaxy (M31), identified using a MeanShift-based algorithm on *HST*'s F275W-band imaging from the PHAT (Dalcanton et al. 2012) and PHAST (Chen et al. 2025) Hubble surveys. These clusters exhibit typical half-light radii of 1–2 pc and strong ultraviolet luminosities indicative of recent massive star formation. Spatial analysis reveals a pronounced north–south asymmetry: clusters in the northern disc show tight associations with giant molecular clouds (GMCs), while southern clusters appear more compact and luminous but less correlated with molecular gas. Two-point correlation functions demonstrate significant clustering of OBC candidates on scales 100 pc and a strong spatial association with GMCs, consistent with hierarchical star formation in dense gas-rich environments. These findings offer new constraints on the early evolution and feedback-driven dispersal of young stellar clusters across galactic discs.

## 1 INTRODUCTION

Star formation is a fundamental process in galactic evolution, intricately linked to the structure and dynamics of the interstellar medium (ISM) (Kennicutt 1998; McKee & Ostriker 2007). Giant molecular clouds (GMCs), as the primary reservoirs of cold, dense molecular gas, serve as the cradles of star formation (Solomon et al. 1987; Heyer & Dame 2015). Within these environments, massive O- and B-type stars often form in compact clusters that exert intense radiative and mechanical feedback on their surroundings (Zinnecker & Yorke 2007; Krumholz 2014). Understanding the spatial association between young stellar clusters and their natal GMCs is thus essential to constraining the efficiency, duration, and environmental dependence of star formation (Lada & Lada 2003; Chevance et al. 2020).

In the Milky Way, associations between young clusters and GMCs have been extensively studied using molecular line and infrared surveys (e.g. Lada & Lada 2003; Moore et al. 2015; Fukui et al. 2021). These studies suggest that many young clusters remain embedded in, or close to, their natal clouds during the first few million years. However, the disruptive effects of stellar feedback – via ionizing radiation, winds, and supernovae – can quickly disperse the surrounding gas, shortening the window for detecting such associations.

Nevertheless, studies within the Milky Way are often limited by line-of-sight confusion and projection effects (Krumholz 2014; Chevance et al. 2020). By contrast, M31 (the Andromeda Galaxy) offers a cleaner observational laboratory for studying these spatial relationships. Its proximity and moderate inclination allow for high-resolution, face-on views of its stellar and gaseous components.

Over the past decade, several large-scale surveys have provided rich, multiwavelength data sets for M31. These include the *Galaxy Evolution Explorer* (*GALEX*; Thilker et al. 2005) UV imaging, the Hubble-based Panchromatic Hubble Andromeda Treasury (PHAT; Dalcanton et al. 2012), and most recently, the Panchromatic Hubble Andromeda Southern Treasury (PHAST; Chen et al. 2025), which extends the PHAT coverage into the southern disc of M31. Alongside the far-infrared Herschel Exploitation of Local Galaxy Andromeda (HELGA; Kirk et al. 2014) programme, these data sets have enabled the construction of extensive catalogues of stellar clusters (e.g. Vansevičius et al. 2009; Kang et al. 2012; Johnson et al. 2015; Lewis et al. 2015) and GMCs (e.g. Rosolowsky 2007; Kirk et al. 2014), laying the groundwork for systematic investigations of star formation across M31's disc (e.g. Grasha et al. 2018; Peltonen et al. 2023).

While previous studies have explored the link between massive stars and their natal molecular environments in nearby galaxies, a systematic census of UV-bright stellar groupings – hereafter referred to as OBCs – across M31, particularly in its southern disc, remains incomplete. Furthermore, the spatial correspondence between these young clusters and GMCs on sub-kiloparsec scales is not yet well constrained. Understanding whether OBCs consistently trace dense molecular gas, or whether stellar feedback rapidly disrupts their natal clouds, remains an open question in extragalactic star formation studies (e.g. Faucher-Giguère, Quataert & Hopkins 2013; Li et al. 2019; Fukui et al. 2021).

In this work, we present a comprehensive catalogue of OBC candidates in M31 based on the *HST* F275W images, with a focus on the previously underexplored southern disc. By cross-matching the clusters with the GMC catalogue from the HELGA survey (Kirk et al. 2014), we perform a statistical analysis of their spatial relationships.

★ E-mails: lgw@bao.ac.cn (GWL), cwtsai@nao.cas.cn (CWT), jingwen@nao.cas.cn (JW)

This paper is organized as follows: Section 2 describes the data sets and sample selection; Section 3 outlines the methodology; Section 4 presents the results; Section 5 discusses the implications; and Section 6 summarizes our conclusions.

## 2 DATA

### 2.1 The panchromatic Hubble Andromeda treasury

The panchromatic Hubble Andromeda treasury (PHAT; Dalcanton et al. 2012) is a *Hubble Space Telescope* Multi-Cycle Treasury programme that imaged approximately one-third of the star-forming disc of M31. The survey covers a contiguous 0.5 deg$^2$ region across 828 orbits, utilizing six broad photometric bands ranging from the ultraviolet (F275W, F336W) to the near-infrared (F110W, F160W), observed with both WFC3 and ACS instruments. This data set resolves millions of stars across galactocentric radii from 0 to 20 kpc.

In the low-density outer disc, the photometry achieves a signal-to-noise ratio (S/N) of 4 at $m_{F275W} = 25.1$ and $m_{F475W} = 27.9$ (Dalcanton et al. 2012). However, crowding in the inner regions reduces the effective depth by up to five magnitudes. Observations followed a two-orbit-per-pointing strategy with Nyquist-sampled imaging in the F475W, F814W, and F160W bands.

PHAT enables reliable measurements of stellar temperatures, bolometric luminosities, and extinctions, with extensive validation of photometric stability and crowding. Density maps of red giant branch (RGB) stars show that the well-known 10 kpc ring hosts not only recent star formation but also a significant overdensity of stars older than 1 Gyr, underlining its role as a long-lived dynamical structure.

### 2.2 The panchromatic Hubble Andromeda southern treasury

The panchromatic Hubble Andromeda southern treasury (PHAST; Chen et al. 2025) extends high-resolution imaging to the southern disc of M31. Covering approximately 0.45 deg$^2$ and reaching out to ∼13 kpc along the southern major axis, PHAST brings the total star-forming disc coverage to roughly two-thirds.

The survey provides deep photometry of over 90 million resolved stars across F275W and F336W bands at UV wavelengths, as well as, F475W and F814W bands at optical wavelengths. Photometric completeness reaches $m_{F475W} \approx 27.7$ in uncrowded regions and ≈ 26.0 near the bulge (Dalcanton et al. 2012). Artificial star tests were conducted to characterize completeness, depth, and uncertainties.

In this work, OBC candidates were initially identified from F275W images due to their sensitivity to hot, young stars. Additional inspection in F336W, F475W, and F814W bands allowed multiband validation of morphology and spatial coherence.

### 2.3 The Herschel exploitation of local galaxy Andromeda

To trace molecular gas, we utilize the catalogue of 326 giant molecular clouds (GMCs) compiled by Kirk et al. (2014) from the Herschel Exploitation of Local Galaxy Andromeda (HELGA) survey. These GMCs were extracted from Herschel far-infrared maps using a hierarchical structure-identification algorithm.

Given Herschel's limited angular resolution, the identified GMCs likely represent cloud complexes rather than individual clouds. Nonetheless, the catalogue provides a uniform and statistically consistent tracer of molecular gas structures across the disc of M31, enabling quantitative assessment of OBC-GMC spatial correlations.

## 3 METHODOLOGY

### 3.1 Image segmentation and grid selection

Detecting extragalactic clusters relies on identifying local overdensities of point sources in photometric images, which manifests as local flux enhancements. Given the challenges of isolating such structures across entire high-resolution images, we divided each brick into smaller grids to facilitate local background estimation and enhance detection sensitivity.

We utilized the F275W band images from the WFC3/UVIS camera, processed through standard *HST* calibration routines (bias correction, flat-fielding, geometric distortion correction, and drizzle resampling). The images have a pixel scale of 0.04 arcsec, corresponding to ∼0.15 pc per pixel at the distance of M31.

The typical diameter of OBCs is several parsecs (Motte, Bontemps & Louvet 2018), translating to tens of pixels on the F275W images. To ensure comprehensive detection while maintaining computational efficiency, we adopted grid sizes ranging from 25 to 150 pixels (3.75–22.5 pc), optimized based on detection completeness and background sensitivity tests.

All PHAT and PHAST imaging data are publicly available via the Mikulski Archive for Space Telescopes (MAST) at https://archive.stsci.edu/hlsp/phat and https://archive.stsci.edu/hlsp/phast.

### 3.2 Identification of cluster centres using the meanshift algorithm

We adopted the MeanShift algorithm (Fukunaga & Hostetler 1975), a non-parametric density gradient estimator, to identify cluster centres. This method has been widely used in various tasks such as pattern recognition and image segmentation since the late 1990s (Comaniciu & Meer 1997; Bradski 1998; Comaniciu & Meer 1999). It is also well-suited to astronomical applications (e.g. Li & Zhao 2009) as it locates local density maxima without requiring pre-defined cluster shapes or numbers.

For each stellar cluster candidate, we defined a circular search region $C$ of radius $R$, with the same value of grid size in Section 3.1, centred at an initial position $P_0 = (x_0, y_0)$. The updated centre $P_1$ is computed iteratively as the weighted centroid:

$$P_1 = \frac{\sum_{i \in C} \omega_i I_i}{\sum_{i \in C} \omega_i}, \qquad (1)$$

where $I_i = (x_i, y_i)$ is the position of pixel $i$ and $\omega_i$ is its flux. The distance of displacement is then:

$$\Delta r = \|P_1 - P_0\|. \qquad (2)$$

If $\Delta r < 1$ pixel, the convergence criterion in our tests, $P_1$ is adopted as the final cluster centre. Otherwise, $P_0$ is updated to $P_1$ and the iteration continues.

### 3.3 Cluster size estimation: full-light and half-light radius

After determining the photometric centre of each candidate cluster, we define the full-light radius ($R_{full}$) using a signal-to-noise ratio (SNR) based method as follows. Within each grid, the standard deviation of the background flux is calculated to estimate the local noise level, and pixels with flux values exceeding $5\sigma$ are considered as signal.

The initial $R_{\text{full}}$ is set as the distance from the cluster centre to the most distant signal pixel. All candidates are then examined visually to correct for false positives or missed detections caused by background fluctuations or complex morphologies. Based on the typical radii of OBCs (e.g. $\sim 3$ pc; Bastian et al. 2012; Motte et al. 2018) and the spatial resolution of *HST* imaging, we adopt a minimum $R_{\text{full}}$ threshold of 3 pixels to ensure sufficient spatial extent.

Within this radius, point sources were initially identified using the DAOStarFinder function from the PHOTUTILS package (Bradley et al. 2024). Foreground and background contaminants were then excluded by cross-matching with Gaia DR3 (Gaia Collaboration 2021; Evans et al. 2023), removing sources inconsistent with M31 membership. All candidates were subsequently visually inspected to ensure reliability, particularly in high-density regions. To exclude isolated extended sources and mitigate contamination from noise or chance stellar alignments, we required that at least five point sources be detected within $R_{\text{full}}$ for a candidate to be classified as an OBC.

The half-light radius ($R_{\text{eff}}$) is another key structural parameter for extragalactic star clusters. Both the King (King 1962) and Elson, Fall and Freeman (EFF) (Elson, Fall & Freeman 1987) models are widely used to describe the surface brightness profiles of star clusters. The King model is suitable for dynamically relaxed, centrally concentrated clusters, while the EFF model is more appropriate for young, loosely bound clusters with extended haloes. Given the diversity in morphology and dynamical age among our candidates, we avoid arbitrarily favouring one model over the other. Instead, we adopt the average of $R_K$ and $R_E$ as a representative estimate of the half-light radius.

$$R_{\text{eff}} = \frac{R_E + R_K}{2}. \qquad (3)$$

This approach, consistent with previous studies (e.g. Vansevičius et al. 2009), provides a robust compromise between the two structural models. The relative differences are modest (median $\sim 13.8$ per cent), and the use of averaged values does not impact the main trends or conclusions.

### 3.4 Photometric measurement

To measure the apparent magnitude of a star cluster, we adopt a procedure broadly inspired by that of Johnson et al. (2015). While the overall methodology is similar, we note that certain implementation details in the original description are not fully specified. Therefore, our actual procedure includes modifications tailored to our data set.

We begin by defining a circular photometric aperture centred on the cluster with radius $R_{\text{full}}$. The local background is estimated from an annular region extending from $1.2\,R_{\text{full}}$ to $3.4\,R_{\text{full}}$.

Aperture photometry is then performed, and the background-subtracted flux within the aperture is denoted as $F_{\text{net}}$. The apparent magnitude $m$ is computed using the instrumental zero-point magnitude $m_{\text{zp}}$.

Because a finite aperture may miss flux from the outer regions of the cluster, particularly in its extended wings, the resulting magnitude tends to be systematically underestimated. An aperture correction is therefore required.

For the photometric measurements for clusters obtained in the PHAT and PHAST surveys, we assume that the cluster surface brightness follows a King profile (King 1962) with a fixed concentration parameter $c = 7$. We numerically compute the cumulative distribution function (CDF) of this profile to estimate the enclosed flux fraction. The ratio of the aperture radius to the total (truncation) radius, $r_{\text{ratio}} = R_{\text{full}}/R_{\text{total}}$, determines the fraction of total flux enclosed, denoted as $f(r_{\text{ratio}})$. The corresponding aperture correction in magnitudes is given by:

$$\Delta m_{\text{apcor}} = -2.5 \log_{10}(f(r_{\text{ratio}})). \qquad (4)$$

The corrected apparent magnitude becomes:

$$m_{\text{corrected}} = m + \Delta m_{\text{apcor}}. \qquad (5)$$

In cases where the total radius is not independently measured, we adopt an approximate scaling of $R_{\text{total}} = 2.5 \times R_{\text{full}}$, based on typical cluster profiles. This approximation effectively compensates for flux lost outside the aperture and improves the accuracy of the magnitude estimates.

To ensure transparency and reproducibility, all computational procedures and parameters used in our implementation are available upon request.

### 3.5 Two-point correlation function

To quantify the spatial clustering properties of OBC candidates and GMCs, we calculate the projected two-point correlation functions using the Landy & Szalay (1993) estimator. The Landy–Szalay estimator is adopted because it is widely used in studies of star cluster distributions in nearby galaxies (e.g. Turner et al. 2022; Peltonen et al. 2023). Specifically, we compute the autocorrelation function $\omega(r)$ and the cross-correlation function $\zeta(r)$ between the two samples.

The autocorrelation function is defined as:

$$\omega(r) = \frac{N_R(N_R-1)}{\text{RR}(r)} \left[ \frac{\text{DD}(r)}{N_D(N_D-1)} - \frac{\text{DR}(r)}{N_D N_R} + \frac{\text{RR}(r)}{N_R(N_R-1)} \right], \qquad (6)$$

where DD($r$), DR($r$), and RR($r$) denote the number of data–data, data–random, and random–random pairs, respectively, at projected separation $r$. $N_D$ and $N_R$ represent the total numbers of data and random points.

For cross-correlation between two different samples (e.g. OBC candidates and GMCs), we use the generalized Landy–Szalay estimator:

$$\zeta(r) = \left( \frac{N_{R1}N_{R2}}{N_{D1}N_{D2}} \cdot \frac{D_1 D_2(r)}{R_1 R_2(r)} \right) - \left( \frac{N_{R1}}{N_{D1}} \cdot \frac{D_1 R_2(r)}{R_1 R_2(r)} \right) - \left( \frac{N_{R2}}{N_{D2}} \cdot \frac{R_1 D_2(r)}{R_1 R_2(r)} \right) + 1, \qquad (7)$$

where $D_1 D_2(r)$ denotes the number of data–data cross-pairs at separation $r$ between the two samples. $D_1 R_2(r)$ and $R_1 D_2(r)$ represent the number of cross-pairs between data in one sample and random points in the other. $R_1 R_2(r)$ is the number of random–random pairs. The total number of data and random points in sample 1 and sample 2 are denoted by $N_{D1}$, $N_{D2}$, $N_{R1}$, and $N_{R2}$, respectively.

Random catalogues were generated by uniformly redistributing positions within the spatial extent of each real data set. Each random catalogue contains ten times as many points as its corresponding data sample, ensuring good statistical convergence in pair counts.

To estimate uncertainties in the correlation functions, we generate 100 independent realizations of random catalogues for each sample. For each realization, the corresponding auto- or cross-correlation function is computed using the same Landy–Szalay formalism. The $1\sigma$ error in each radial bin is then derived from the standard deviation of the correlation values across these 100 trials. This Monte Carlo

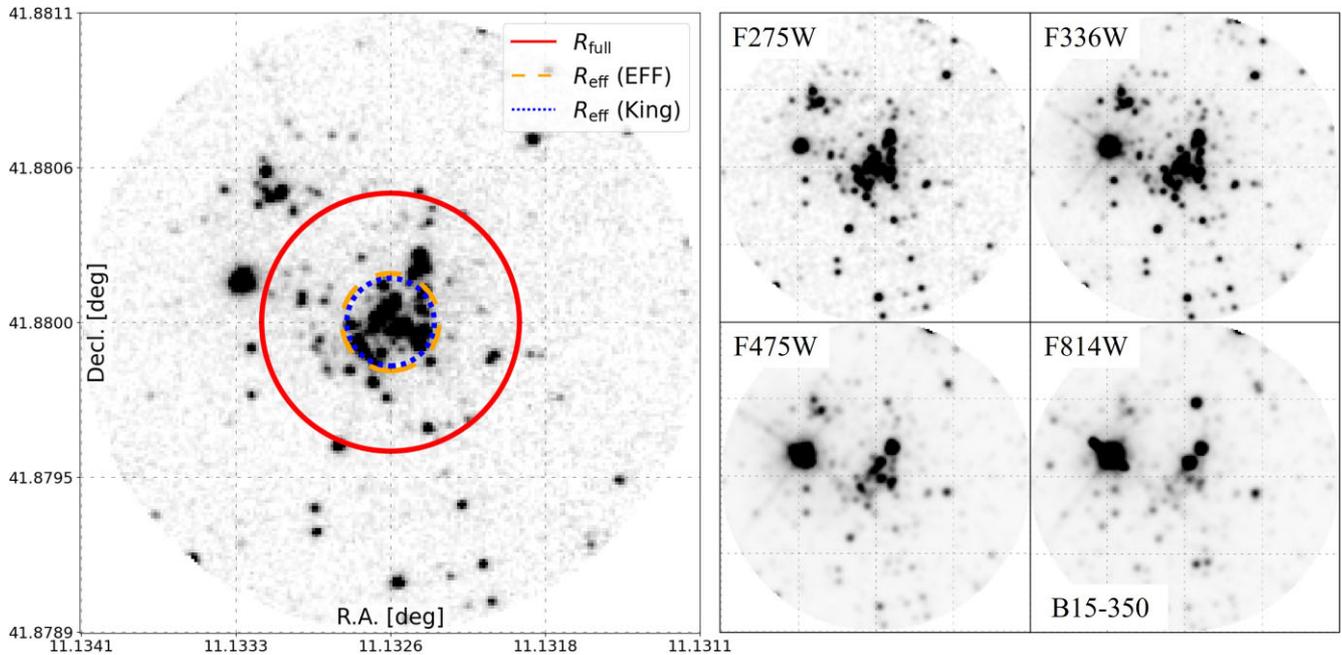

**Figure 1.** Multiband imaging of a representative OBC candidate. Panel (a) shows the F275W image centred on the cluster. The solid circle indicates the full-light radius. The dashed and dotted circles denote the half-light radii derived from the EFF and King profile fits, respectively. Panel (b) displays comparison images in the F336W, F475W, and F814W bands, highlighting the enhanced flux concentration in the ultraviolet (F275W), which is characteristic of young, luminous OBC candidates.

**Table 1.** The catalogue of OBC candidates in M31.

| YLID | Brick | R.A. (Deg) | Decl. (Deg) | $R_{\text{full}}$ ($''$) | $R_{\text{eff}}$ (pc) | $m_{\text{F275W}}$ |
|---|---|---|---|---|---|---|
| 1 | b01 | 10.838138 | 41.214752 | 1.862 | 3.93 | 20.3 |
| 2 | b01 | 10.817428 | 41.242983 | 1.426 | 2.71 | 21.8 |
| 3 | b01 | 10.810671 | 41.266607 | 1.585 | 2.40 | 20.5 |
| 4 | b01 | 10.607419 | 41.303992 | 1.030 | 0.43 | 20.5 |
| ⋮ | ⋮ | ⋮ | ⋮ | ⋮ | ⋮ | ⋮ |

*Note.* Column 'YLID' provides a unique identifier for each OBC candidate. 'Brick' indicates the PHAT/PHAST survey region in which the candidate is located. $R_{\text{full}}$ and $R_{\text{eff}}$ denote the full-light radius (in arcseconds) and half-light radius (in parsecs), respectively. $m_{\text{F275W}}$ gives the apparent magnitude in the HST F275W filter. The full version of this table is available at https://nadc.china-vo.org/res/r101512/.

approach captures statistical variance arising from finite-sample effects and random catalogue fluctuations.

## 4 RESULTS

After MeanShift selection and subsequent visual inspection, we identified a total of 578 OBC candidates. Fig. 1(a) shows a representative example of one such candidate. The dashed circle marks the full-light radius obtained via the MeanShift and SNR methods, while the dashed and the dotted circles indicate the half-light radii fitted using the EFF and King profiles, respectively.

To further assess the cluster's nature, we examined its photometric appearance in additional *HST* bands (F336W, F475W, and F814W). As shown in Fig. 1(b), the cluster is clearly identifiable in the F275W and F336W bands, while it appears more diffuse in the F475W and F814W images. The morphology in the near-ultraviolet (especially F275W) is consistent with expectations for a young, luminous OBC dominated by early-type stars.

All OBC candidates are compiled into a catalogue, a subset of which is presented in Table 1. Each entry lists a unique identifier (YLID), the corresponding PHAT/PHAST survey brick, equatorial coordinates (J2000), and structural parameters including the full-light radius ($R_{\text{full}}$, in arcseconds) and half-light radius ($R_{\text{eff}}$, in parsecs). The apparent magnitude in the F275W filter, $m_{\text{F275W}}$, is also provided.

As summarized in Table 1, the candidates span multiple survey bricks and generally exhibit compact morphologies and strong ultraviolet flux, in agreement with expectations for young, massive stellar clusters composed of OB-type stars.

### 4.1 Spatial distribution of OBC candidates in M31

To investigate the spatial distribution of OBC candidates in M31, we overlaid their positions onto the far-infrared dust emission map from Draine et al. (2013). Fig. 2 presents this spatial distribution, with OBC candidates shown as points against a grey-scale background tracing the dust mass column density. The map highlights the prominent star-forming ring and spiral structures of M31. Dashed ellipses denote projected galactocentric distances, while rectangular grids delineate the PHAT (b01–b23) and PHAST (b24–b36) survey footprints.

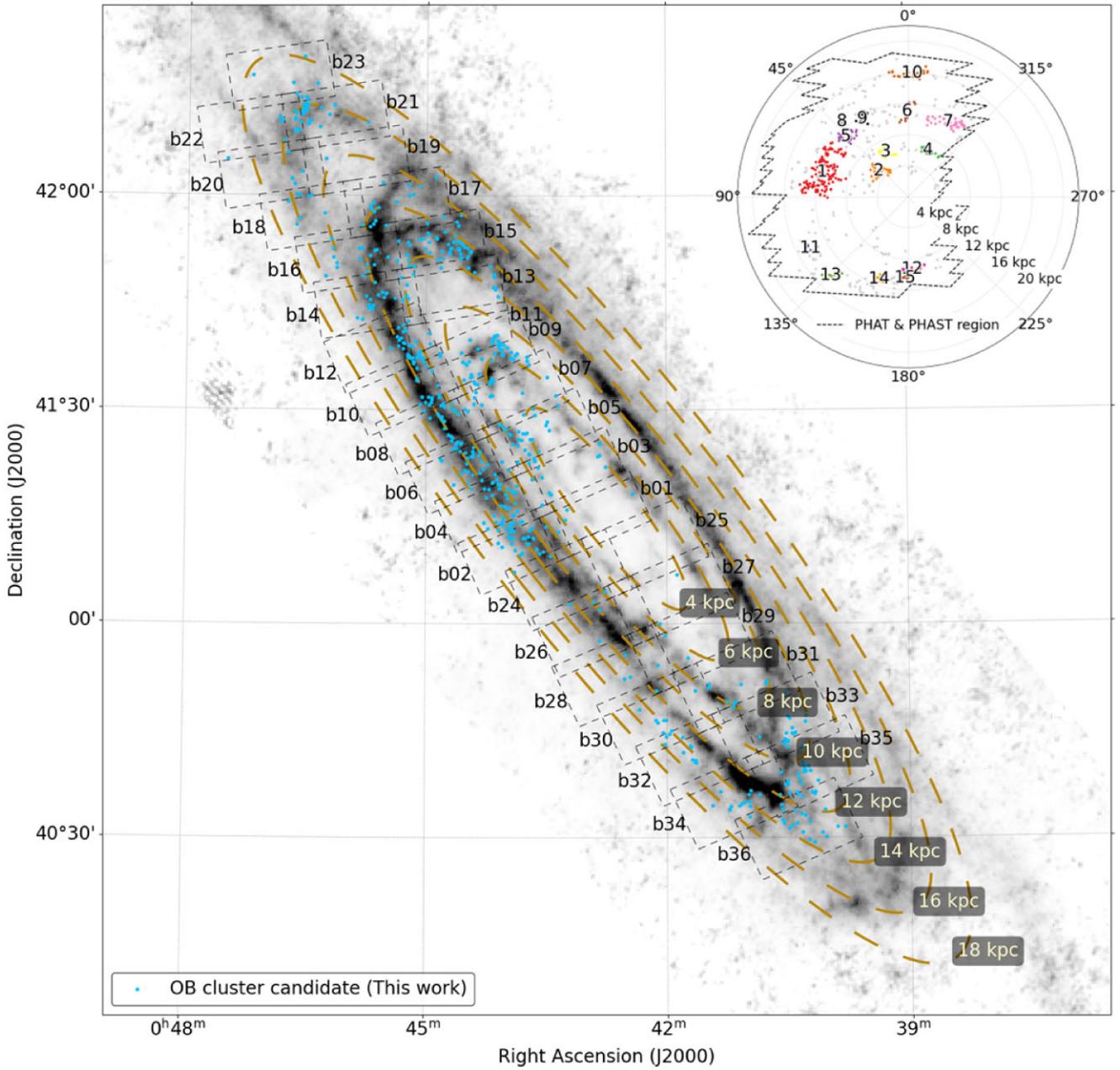

**Figure 2.** Spatial distribution and clustering analysis of OB star cluster candidates in M31. The main panel shows the positions of OBC candidates, marked as solid circular points, overlaid on a far-infrared dust emission map based on publicly available data from Draine et al. (2013), where the grey-scale background traces the dust mass column density, highlighting the prominent ring and spiral features of M31. Dashed ellipses indicate projected galactocentric distances in 2 kpc intervals, assuming an inclination angle of 77° and a position angle of 38.1°. Rectangular grids delineate the PHAT (b01–b23) and PHAST (b24–b36) survey footprints. The inset panel presents the deprojected spatial distribution of OBC candidates in polar coordinates, with groups identified via a minimum spanning tree (MST) analysis.

The majority of OBC candidates in the northern disc exhibit a strong spatial correlation with the well-known ∼10 kpc star-forming ring, aligning closely with regions of high dust column density. This correspondence suggests a physical association between recent massive star formation and the dense interstellar medium. In contrast, OBC candidates in the southern disc are more dispersed, often located along the outer edge of the ring or in offset regions, avoiding peak dust emission zones.

Within the inner disc (6–10 kpc), OBC candidates are distributed across both northern and southern regions. Although the northern sectors show a higher cluster density, the southern presence indicates that massive star formation persists, albeit at a lower level. No OBC candidates are found within the central bulge ($R < 2$ kpc), consistent with the well-established quenching of recent star formation in M31's bulge-dominated nucleus.

A small number of OBC candidates are detected in the outer disc ($R > 14$ kpc), predominantly in the northern region. The apparent absence of such clusters in the south likely reflects the limited spatial extent of the PHAST survey, which covers the southern disc only out to ∼13 kpc. This asymmetry underscores the influence of observational footprint on the perceived distribution of young stellar populations.

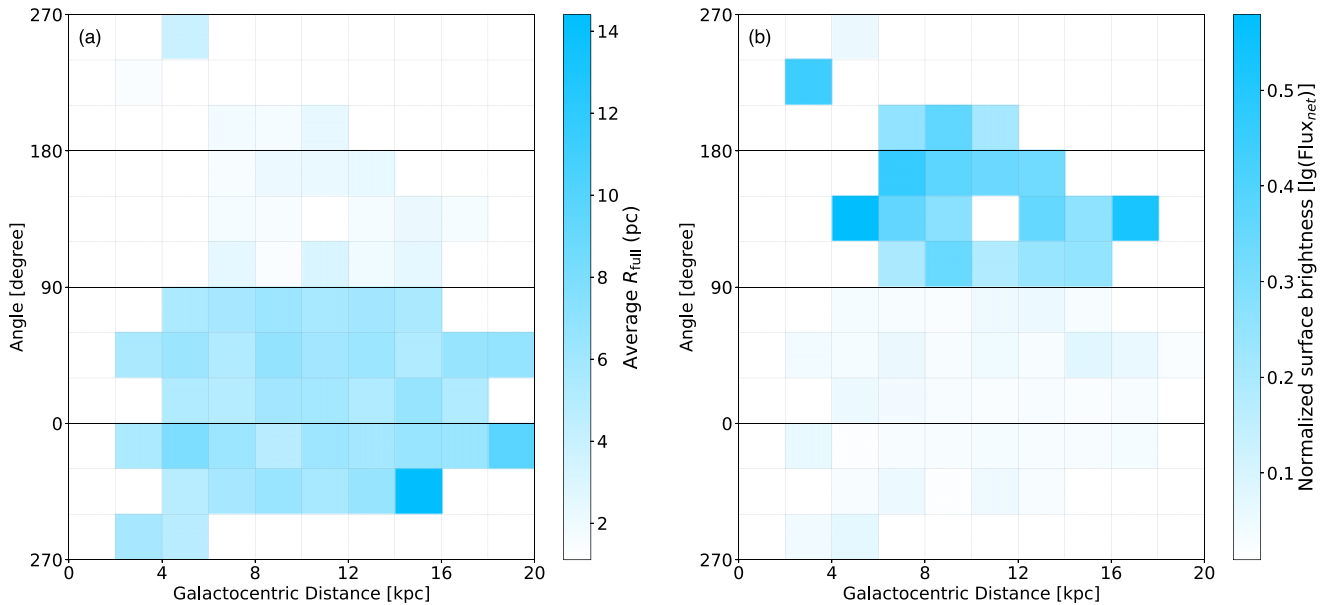

**Figure 3.** 2D polar maps of OBC candidate properties in M31. Panel (a) shows the distribution of mean full-light radius ($R_{full}$), while panel (b) presents the normalized surface brightness. Both maps use radial bins of 2 kpc and azimuthal bins of 30°.

The inset panel of Fig. 2 presents the deprojected spatial distribution of OBC candidates in polar coordinates, where azimuthal angles are measured counterclockwise from the projected northern major axis (0°), and galactocentric radii are indicated by concentric circles.

This polar representation clearly illustrates a pronounced hemispheric asymmetry: 457 OBC candidates (∼79 per cent) are located in the northern half of the disc (0°–90°, and 270°–360°), while only 121 (∼21 per cent) lie in the southern half (90°–270°). This discrepancy may be attributed to two primary factors. First, the observational footprint is inherently asymmetric – the PHAST survey covers the southern disc only out to a projected radius of ∼13 kpc in radius from the centre of M31, whereas the northern coverage extends to ∼16 kpc. Secondly, the southern disc within 12 kpc exhibits heavier dust obscuration (e.g. Draine et al. 2013; Ade et al. 2015), which may further attenuate the ultraviolet emission from young stellar populations, thereby concealing some OBC candidates from detection. In addition, the intrinsic stellar density in the southern half is lower than in the north, as evident from the processed stellar source catalogue, with particularly sparse coverage in bricks 24 and 26. We will revisit and expand on these contributing factors in Section 5.1.

Clusters are number-coded by group, as identified through a minimum spanning tree (MST) analysis using a linking length of 650 pc and a minimum group size of seven OB candidates. The parameters of 650 pc and seven samples were selected after exploring a range of values. These settings yielded clustering results that were consistent with the prominent spatial associations visible in the OB star distribution, providing a physically meaningful partition of the data. A total of 15 groups are identified, revealing large-scale spatial associations and coherent clustering across the disc.

Compared to the projected map, the polar representation provides a clearer and more quantitative view of spatial clustering. The distribution is strongly anisotropic, with the majority of clusters concentrated in the northern sector. Several MST groups (e.g. Groups 1, 5, 6, 7, 8, and 9) are tightly aligned along the 10 kpc ring, exhibiting high azimuthal coherence. Groups 2, 3, and 4 lie interior to the ring (6–8 kpc), while Group 10 extends outward (12–14 kpc), tracing the outer boundary of the star-forming disc.

In the southern sector, OBC candidate groups appear more sparsely distributed with greater azimuthal dispersion. While Groups 12 and 15 partially follow the southern segment of the 10 kpc ring, the remaining groups are more loosely arranged along curved trajectories, forming an open, spiral-like configuration extending from 8 to 16 kpc in radius.

### 4.2 Size and surface brightness of OBC candidates in M31

To characterize the photometric properties of OBC candidates in M31, we constructed 2D polar histograms of key structural parameters, binned by galactocentric radius, and azimuthal angle (Fig. 3). Each map employs radial intervals of 2 kpc and azimuthal segments of 30°.

Panel (a) of Fig. 3 displays the spatial distribution of the mean full-light radius ($R_{full}$) of OBC candidates. The average cluster size exhibits a marked dependence on azimuthal angle, while its variation with radius is comparatively modest. Clusters located in the northern half of the disc generally possess larger sizes, frequently exceeding 8–10 pc, with local maxima reaching up to ∼12 pc. In contrast, those in the southern half exhibit smaller sizes, typically in the range of 3–6 pc. Radially, the distribution remains relatively uniform, with no significant monotonic trend observed across the 0–20 kpc range.

Panel (b) of Fig. 3 presents the corresponding distribution of normalized surface brightness. For each OBC candidate, the surface brightness was computed as the net F275W flux (in units of electrons per second) divided by the area enclosed within the cluster's full-light radius, effectively yielding a flux surface density that reflects both luminosity and spatial compactness. The resulting values were converted to logarithmic scale (base-10), then linearly rescaled to the [0,1] interval for visualization, as labelled 'lg($Flux_{net}/R_{full}^2$)' in Fig. 3(b). Each bin represents the mean of these normalized values, with bluer tones corresponding to more luminous and compact clusters. Notably, the angular dependence of surface brightness shows a reversed pattern relative to $R_{full}$: clusters in the southern half of the disc are systematically brighter (normalized values frequently > 0.5), while northern clusters tend to exhibit lower surface brightness on average.

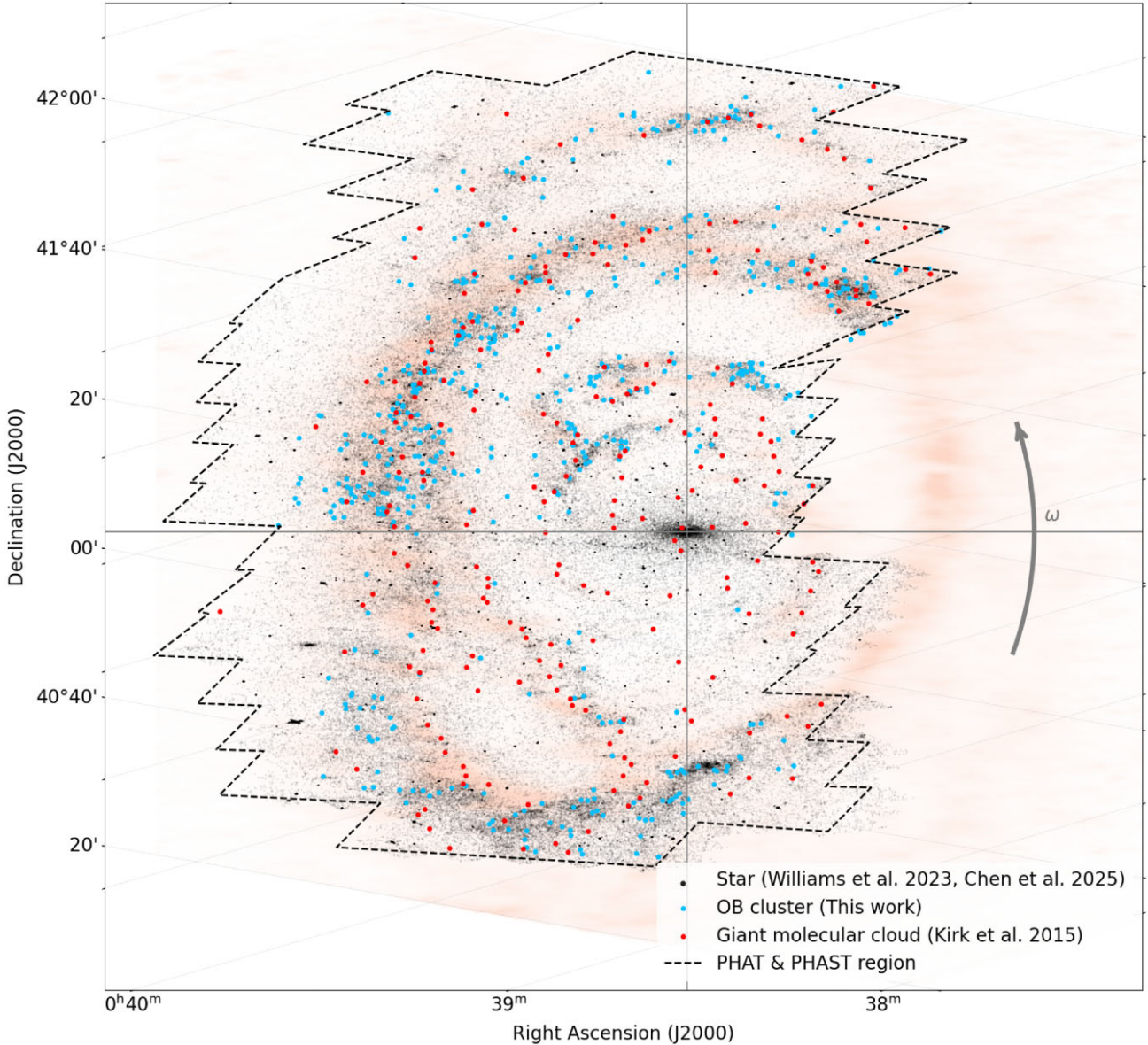

**Figure 4.** Spatial distribution of dust (orange background), bright stars ($m_{F475W} \leq 23$, black dots), giant molecular clouds (red dots), and OBC candidates (blue dots) across M31, shown in a galactocentric deprojected polar coordinate frame.

This angular asymmetry in size and brightness reflects previously reported north–south differences in the stellar and gaseous components of M31, often attributed to large-scale dynamical effects such as disc warping, tidal perturbations, or past interactions with satellite galaxies (e.g. Williams 2003; Tabatabaei & Berkhuijsen 2010; Dalcanton et al. 2012; Lewis et al. 2015). Such disturbances may hinder the formation of bound OBC candidates by disrupting the spatial coherence of star-forming regions. The observed patterns thus underscore the complex interplay between gas dynamics, dust structure, and stellar feedback in shaping the recent star formation activity across M31.

## 5 DISCUSSION

### 5.1 Association between OBC candidates and bright stars in M31

In addition to examining the intrinsic properties of OBC candidates, we investigate their spatial correlation with bright stars (Dalcanton et al. 2012; Chen et al. 2025) ($m_{F475W} \leq 23$) and GMCs in M31 (Kirk et al. 2014). Fig. 4, which presents the deprojected view of M31 in galactocentric coordinates, presents a composite multiwavelength map of M31, incorporating dust emission (orange background), optical stellar sources (black dots), GMCs (red dots), and the newly identified OBC candidates (blue dots).

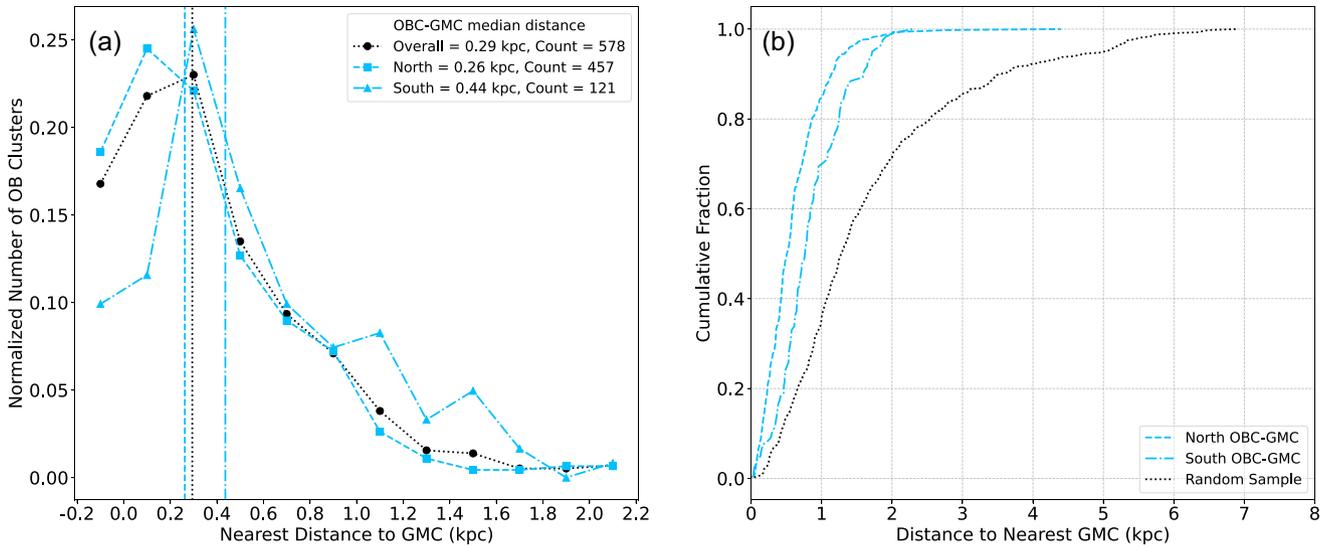

**Figure 5.** Panel (a): Line plot of the projected distances from each OBC candidate to its nearest GMC, defined as the centre-to-centre distance minus the GMC radius. The x-axis shows the nearest distance to a GMC in kiloparsecs (kpc), and the y-axis indicates the normalized number of OBC candidates, i.e. the fraction of the total sample in each bin. The northern region, shown as a dashed line with square markers, includes 457 OBC candidates and has a median separation of 0.26 kpc. The southern region, shown as a dash-dotted line with triangle markers, includes 121 candidates with a median separation of 0.44 kpc. The full sample of 578 OBC candidates is shown as a dotted line with solid circle markers, with a median separation of 0.29 kpc. Panel (b): Cumulative distribution functions (CDFs) of the nearest projected separations between OBC candidates and GMCs. The x-axis indicates distance to the nearest GMC (kpc), and the y-axis shows the cumulative fraction of OBC candidates. The dashed and dash-dotted lines represent the distributions for the northern and southern regions, respectively. The dotted line corresponds to a synthetic sample with randomly distributed GMCs.

To ensure photometric reliability, stars brighter than 23 mag in the F475W filter were selected instead of those in the F275W band, owing to the higher signal-to-noise ratio in the optical regime (Williams et al. 2023; Chen et al. 2025). Two prominent spiral arms are readily discernible – one in the northern disc and one in the southern – consistent with previous observations (e.g. Arp 1964; Simien et al. 1978). Additionally, multiple minor arm-like structures are evident in the northern half of the disc, suggesting a more complex and asymmetric morphology than that expected for classical grand-design spirals. By contrast, the southern disc appears more fragmented, with the primary stellar arm breaking into three or more disconnected segments, in agreement with earlier findings (e.g. Gordon et al. 2006; Tenjes et al. 2017).

A pronounced spatial alignment is observed between OBC candidates and the bright stellar components, particularly along the spiral features. This spatial association supports the scenario in which OBC formation is regulated by spiral arm dynamics, likely via gas compression induced by spiral density waves. Similar spatial correlations between young stellar populations and spiral arms have been reported in the Milky Way and other nearby disc galaxies (e.g. Egusa et al. 2009; Portegies Zwart, McMillan & Gieles 2010; Leroy et al. 2013), highlighting the key role of spiral density enhancements in triggering massive star formation.

Notably, a marked decline in the surface density of bright stellar sources is observed in the southwestern disc of M31. This stellar paucity results in an apparent absence of OBC candidates in these areas. The lower surface density of stars may be still caused by stronger dust extinction, as mentioned in Section 4.1 supported by Draine et al. (2013) and Ade et al. (2015). In addition, the stellar density in this region may genuinely be lower, potentially due to a quiescent star formation history influenced by past interactions or satellite accretion events. (e.g. D'Souza & Bell 2018; Hammer et al. 2018).

These spatial alignments between OBC candidates and spiral structures motivate a more quantitative assessment of their associations with the molecular gas reservoirs that fuel star formation.

### 5.2 Association of OBC candidates with giant molecular clouds in M31

In addition to the asymmetry observed between OBC candidates and dust, a similarly pronounced north–south difference is found in their spatial relationship with GMCs. Both the dust emission map (Draine et al. 2013) and the GMC catalogue (Kirk et al. 2014) are independently used to evaluate spatial correlations with OBC candidates. Although they are derived from different observational pipelines and methodologies, this does not affect our analysis, as we do not attempt a direct comparison between dust and GMCs.

In the northern disc of M31, OBC candidates are closely aligned with GMCs, frequently interleaved along the same segments of the star-forming ring – consistent with sites of ongoing massive star formation. In contrast, the southern quadrants display a systematic offset: OBC candidates are preferentially located outside the dust ring, whereas GMCs tend to lie along its inner edge. This spatial separation is particularly evident in MST groups 11 and 15.

Fig. 5 illustrates the spatial correlation between OBC candidates and their nearest GMCs, separated into northern and southern regions of M31, based on a deprojected, face-on view in polar coordinates. The projected separation is measured from the centre of each OBC candidate to the centre of the closest GMC, corrected for GMC radius.

Panel (a) presents the normalized histogram of these nearest distances. The x-axis represents the projected separation to the nearest GMC in kiloparsecs (kpc), while the y-axis denotes the normalized number of OBC candidates, i.e. the fraction of the total sample falling within each bin. A clear difference emerges between the two regions: northern OBCs (457 objects) show a strongly peaked distribution with a median separation of 0.26 kpc, indicating strong clustering around GMCs. In contrast, the southern sample (121 objects) exhibits a broader and flatter distribution with a higher median of 0.44 kpc, and many candidates located beyond 0.4 kpc, suggesting a weaker spatial association.

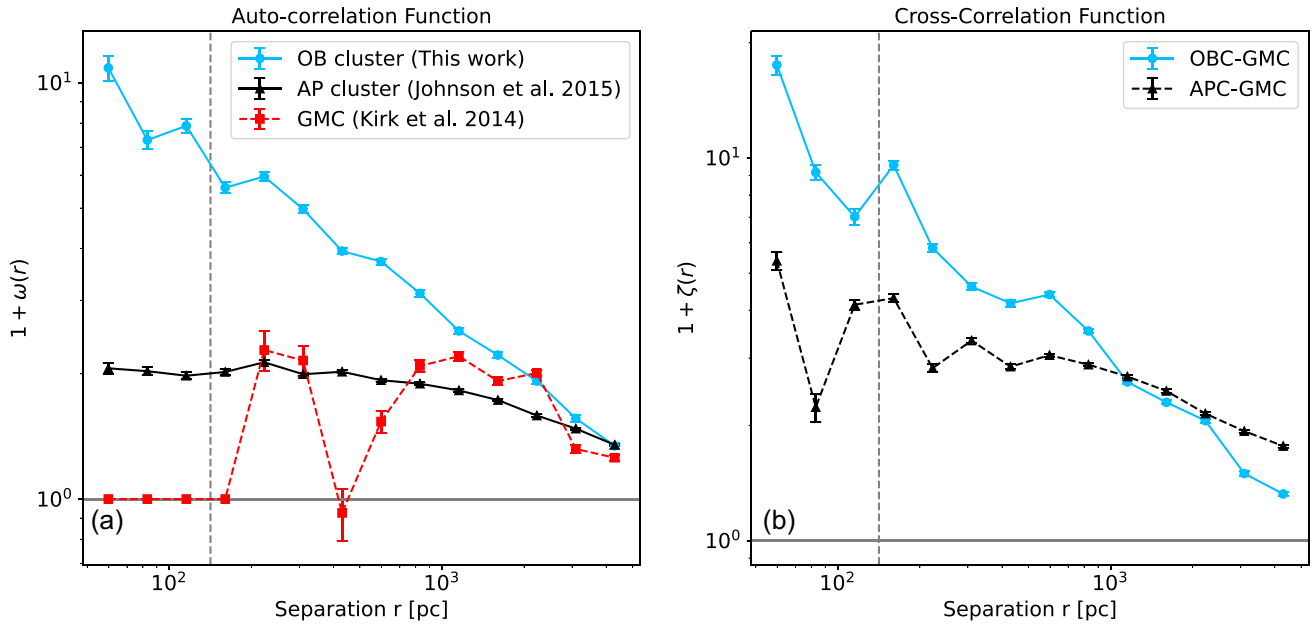

**Figure 6.** Two-point correlation analysis of OBC candidates and GMCs in M31, with $1\sigma$ standard deviation error bars shown. Panel (a): Autocorrelation functions $1 + \omega(r)$ for OBC candidates (solid line with circle markers), Andromeda Project (AP) clusters (solid line with triangle markers), and GMCs (dashed line with square markers), plotted as a function of projected separation $r$ (log–log scale). The vertical dashed line indicates the average GMC radius ($\langle R_{\rm GMC} \rangle = 142$ pc). Panel (b): Cross-correlation functions $1 + \zeta(r)$ between OBC candidates and GMCs (solid line with circle markers), and between AP clusters and GMCs (dashed line with triangle markers). The OBC-GMC correlation peaks at scales comparable to typical GMC sizes, indicating a robust physical association.

Although the median separations of 0.26 and 0.44 kpc may appear numerically close, the relative difference exceeds 40 per cent, marking a substantial spatial contrast. More importantly, this suggests a qualitatively different degree of coupling between the OBCs and GMCs in the two regions. This point will be revisited in the context of age distribution analysis in Section 5.3.

Panel (b) compares the cumulative distribution functions (CDFs) of the OBC-GMC separations in both regions with that of a synthetic, randomly positioned OBC population. Both observed distributions rise significantly more steeply than the random expectation, especially within the first ∼2 kpc. The northern sample exhibits the sharpest rise, consistent with a stronger physical association. These results support a scenario in which OBCs preferentially form in close proximity to GMCs, with the effect being more pronounced in the northern half of M31, in line with theoretical expectations and previous studies (Grasha et al. 2018; Turner et al. 2022).

Fig. 6 presents the two-point correlation functions between OBC candidates and GMCs, incorporating $1\sigma$ Poisson uncertainties to assess statistical significance.

Panel (a) shows the autocorrelation functions $1 + \omega(r)$ for OBC candidates (solid line with circle markers, this work), Andromeda Project (AP) clusters (Johnson et al. 2015) from the PHAT catalogue (solid line with triangle markers), and GMCs identified by Kirk et al. (2014) (dashed line with square markers). OBC candidates exhibit a significantly enhanced clustering amplitude at small scales ($r \lesssim 200$ pc), with $1 + \omega(r) > 10$ at $r < 100$ pc. This steep rise towards smaller separations remains significant above the $1\sigma$ error level, confirming a hierarchical and scale-free spatial distribution consistent with recent *in-situ* formation. In contrast, AP clusters display a flatter correlation profile with larger uncertainties, consistent with spatial diffusion over longer dynamical time-scales. GMCs show only mild clustering at small scales, which may reflect a mix of cloud evolutionary stages or catalogue resolution effects.

Panel (b) plots the cross-correlation functions $1 + \zeta(r)$ between OBC candidates and GMCs (solid line with circle markers), and AP clusters and GMCs (dashed line with triangle markers), again including $1\sigma$ error bars. A strong OB-GMC correlation is observed at $r \sim 100$–150 pc, peaking near the mean GMC radius. This statistically significant excess supports a robust spatial association between OBC candidates and their natal clouds. The weaker, flatter APC-GMC cross-correlation suggests that older clusters have dynamically decoupled from their birth environments.

### 5.3 Age estimation of OBC candidates

The OBC candidates in our sample are inferred to be relatively young, with characteristic ages of 15–20 Myr. Although supported by several observational and theoretical lines of evidence, these estimates are indirect and should be regarded as indicative, not definitive.

First, the OBC candidates exhibit strong ultraviolet (UV) emission (Fig. 1b), which is primarily attributed to hot, massive OB-type stars. Older stellar components, if present, contribute negligibly to the UV flux due to their lower intrinsic luminosities and the *HST* filter throughput. This makes UV brightness a robust, though indirect, tracer of recent massive star formation.

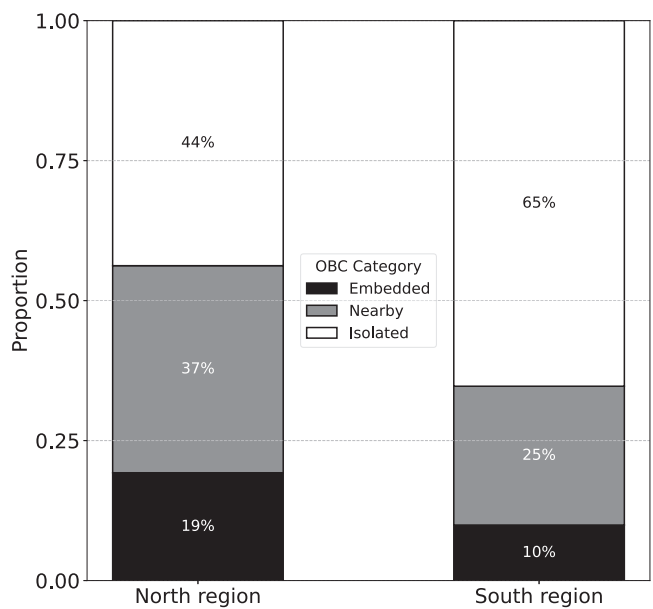

**Figure 7.** Fractional distribution of OB cluster (OBC) categories in the northern (0°–180°) and southern (180°–360°) regions of M31, classified according to their projected distance from the nearest GMC edge. OBCs located within a GMC are defined as 'embedded' candidates (black), those within 300 pc of the nearest GMC edge are labelled as 'nearby' candidates (grey), and the rest are classified as 'isolated' candidates (white). In the northern region, 19 per cent of OBCs are embedded, 37 per cent are nearby, and 44 per cent are isolated; in the southern region, the corresponding fractions are 10 per cent, 25 per cent, and 65 per cent, respectively.

Secondly, based on the statistics of OBC-GMC centre-to-edge (C2E) distances (Fig. 5a), we classify OBC candidates as 'embedded' if located within their nearest GMCs, and as 'nearby' if their projected separation from the GMC edge is less than or equal to 300 pc. The remaining OBCs are classified as 'isolated'. Observational studies in the Milky Way and nearby galaxies suggest that young clusters typically drift from their natal GMCs at velocities of a few to $\sim 20$ km s$^{-1}$ (Dobbs, Bonnell & Pringle 2006; Grasha et al. 2018; Peltonen et al. 2023). Assuming similar kinematic conditions in M31, the 300 pc threshold corresponds to a drift time-scale of $\sim 15$ Myr. This suggests that 'nearby' clusters could be as old as $\sim 15$ Myr, and that the 'isolated' population-comprising nearly half of the OBCs-may be even older, depending on their drift history.

As shown in Fig. 7, the spatial distributions of OBCs differ markedly between regions. In the northern half of M31, 56 per cent of OBC candidates are either embedded within or located near GMCs, indicating a strong spatial association with the molecular gas. In contrast, 65 per cent of southern OBCs are classified as isolated, suggesting weaker or more disrupted coupling between clusters and GMCs. The higher fraction of embedded (19 per cent) and nearby (37 per cent) clusters in the north supports the scenario in which recent or ongoing star formation remains more strongly tied to the GMC population.

Thirdly, although isolated clusters are generally expected to be older under the drift scenario, many – particularly those in MST groups 11 and 15 (Fig. 3b) – exhibit high luminosities, indicative of recent star formation activity. We therefore propose that these OBC candidates may be even younger than embedded or nearby clusters, having rapidly dispersed or disrupted their natal molecular material through strong radiative and mechanical feedback. This interpretation is supported by both simulations and theoretical models of cluster-cloud interactions (Faucher-Giguère et al. 2013; Li et al. 2019; Fukui et al. 2021). However, without direct age constraints from, e.g. SED fitting, we cannot rule out alternative explanations such as projection effects or stochastic sampling of the upper IMF.

Finally, the spatial cross-correlation function between OBC candidates and GMCs (Fig. 6b) declines steeply with increasing separation, in contrast to the much shallower decline observed for AP cluster-GMC correlations. This stronger spatial association is consistent with results from other nearby galaxies, where clusters younger than $\sim 10$ Myr display similarly tight correlations with molecular gas structures (e.g. Turner et al. 2022; Peltonen et al. 2023).

These age estimates are further supported by previous studies using PHAT photometry in M31, which found that clusters with similar spatial and photometric properties are typically younger than 20 Myr (e.g. Beerman et al. 2012; Johnson et al. 2015). While precise CMD fitting is limited in extragalactic contexts due to crowding and depth constraints, the convergence of spatial, photometric, and feedback-based age indicators lends credibility to our classification. None the less, we emphasize that our analysis does not directly measure cluster ages; definitive age constraints will require spectral or photometric modelling.

## 6 CONCLUSION

We present a catalogue of 578 OBC candidates in M31, identified via multiwavelength photometry and MeanShift clustering. These clusters are compact ($R_{\rm eff} \sim 1$–$2$ pc), UV-bright, and predominantly located along the 10 kpc star-forming ring.

A strong spatial association is observed between OBC candidates and GMCs, with median cluster-cloud separations of $\sim 0.3$ kpc. Two-point correlation functions reveal enhanced clustering of OBC candidates on scales $\lesssim 100$ pc, consistent with a hierarchical distribution inherited from turbulent, gas-rich birth environments. These spatial correlations, combined with proximity to molecular gas and UV luminosities, indicate that the majority of OBC candidates are younger than $\sim 20$ Myr.

We find systematic north–south asymmetries in OBC candidate morphology and cloud association: southern clusters appear more compact and luminous, yet less tightly linked to GMCs. This may reflect accelerated gas dispersal by stellar feedback or differences in local star formation conditions across the disc.

Our results place observational constraints on the time-scales of massive cluster formation, early evolution, and cloud disruption. The methods developed here provide a scalable framework for identifying and characterizing young stellar clusters in other nearby galaxies using upcoming high-resolution surveys.

Future observational advances will enable more precise characterization of the OBC candidates identified here. In particular, upcoming facilities such as the Chinese Space Station Telescope (CSST) will provide wide-field imaging in the ultraviolet and optical bands with photometric depth and spatial resolution approaching those of *HST*. Although its spatial resolution is somewhat lower, its large field of view will enable efficient surveys of extended regions in M31, allowing more comprehensive studies of the identified cluster candidates. These stellar population analyses will offer an independent cross-check of our statistical clustering-based age inference and help refine our understanding of massive cluster formation and evolution in galactic discs.

## ACKNOWLEDGEMENTS

This work was supported by the National Natural Science Foundation of China (NSFC) under grants No. 12588202, No. 12041302, and No. 12073038, and by the National Key R&D Program of China under grant No. 2023YFA1608004.

Data resources were provided by the China National Astronomical Data Center (NADC) and the Chinese Virtual Observatory (China-VO). This research also benefited from the Astronomical Big Data Joint Research Center, co-founded by the National Astronomical Observatories, Chinese Academy of Sciences, and Alibaba Cloud.

## DATA AVAILABILITY

The cluster candidate catalogue and associated photometric data underlying this article are available at https://nadc.china-vo.org/res/r101512/. Additional processed files and scripts used in this study are available upon reasonable request from the corresponding author.

This paper has been typeset from a TeX/LaTeX file prepared by the author.